\documentclass[prc,superscriptaddress,noshowpacs,unsortedaddress,twocolumn,showpacs,preprintnumbers,amsmath,amssymb]{revtex4-1}

\usepackage[dvipdfmx]{graphicx}
\usepackage{amsmath,amssymb,times}
\usepackage{color}
\usepackage{ulem}
\usepackage{bm}
\usepackage{here}

\def\gsim{~\,\makebox(1,1){$\stackrel{>}{\widetilde{}}$}\,~}
\def\lsim{~\,\makebox(1,1){$\stackrel{<}{\widetilde{}}$}\,~}

\newcommand{\beq}{\begin{equation}}
\newcommand{\eeq}{\end{equation}}
\newcommand{\bea}{\begin{eqnarray}}
\newcommand{\eea}{\end{eqnarray}}

\newcommand{\bfi}[1]{\mbox{\boldmath $#1$}}

\newcommand{\vK}{{\bfi K}}

\newcommand{\vs}{{\bfi s}}

\newcommand{\vrr}{{\bfi r}}
\newcommand{\vR}{{\bfi R}}

\def\a{\alpha}

\begin{document}

%\preprint{}

\title{Skin values of $^{208}$Pb and $^{48}$Ca determined from reaction cross sections 
}



\author{Tomotsugu~Wakasa}
%\email[]{wakasa@phys.kyushu-u.ac.jp} 

\author{Shingo~Tagami}
%\email[]{sh.tagami@gmail.com}

\author{Masanobu Yahiro}
\email[]{orion093g@gmail.com}
\affiliation{Department of Physics, Kyushu University, Fukuoka 819-0395, Japan} 

\begin{abstract}
\begin{description}
\item[Background]
The PREX and the CREX group reported skin values, 
$r_{\rm skin}^{208}({\rm PREX2}) = 0.283\pm 0.071\,{\rm fm}$ and   
$r_{\rm skin}^{48}({\rm CREX})=0.121 \pm 0.026\ {\rm (exp)} \pm 0.024\ {\rm (model)}$~fm, respectively.  
Using the Love-Franey (LF) $t$-matrix folding model with the neutron and proton densities scaled to the neutron radius $r_{\rm n}^{208}({\rm PREX2})$ and the proton radius of  the electron scattering, we found that the reaction cross sections 
$\sigma_R$ reproduce the data for p+ $^{208}$Pb scattering 
at $E_{\rm lab} = 534.1, 549, 806$~MeV.  
 Zenihiro {\it el al. } deduce neutron radii $r_{\rm n}^{48,40}({\rm exp})$ from the angular
distributions of the cross sections and analyzing powers of proton elastic scattering, 
whereas we determine  matter radius $r_{\rm m}^{\rm 40}({\rm exp})=3.361 \pm 0.075$~fm from measured $\sigma_R$ 
for $^{4}$He+ $^{40}$Ca scattering. 
\item[Aim]
Our first aim is to determine $r_{\rm skin}^{\rm 208}$ from measured $\sigma_R$  of p+$^{208}$Pb scattering at $E_{\rm lab} = 534.1, 549, 806$~MeV by using the Love-Franey (LF) $t$-matrix folding 
model. 
Our second aim is to determine  $r_{\rm skin}^{\rm 48}$ from the $r_{\rm m}^{\rm 40}({\rm exp})$ and 
the difference   $\Delta \equiv r_{\rm m}^{\rm 48}({\rm exp})-r_{\rm m}^{\rm 40}({\rm exp})$ that is evaluated from 
the $r_{\rm n}^{48,40}({\rm exp})$ and 
the $r_{\rm p}^{48,40}({\rm exp})$ calculated with  
 the isotope shift method based on the electron scattering.   
\item[Method and results]
For the first aim, we use the  Love-Franey $t$-matrix model with the  densities scaled from the D1S-GHFB+AMP neutron density, where D1S-GHFB+AMP stands for D1S Gogny HFB (GHFB) with 
the angular momentum  projection (AMP). 
The D1M-GHFB+AMP is also used to estimate a theoretical error. 
The resulting skin values are $r_{\rm skin}^{\rm 208}= 0.324 \pm 0.047$~fm for D1S and 
$r_{\rm skin}^{\rm 208}({\rm exp})=0.333 \pm 0.047 $~fm for D1M.
The difference $\Delta=0.109$~fm and 
$r_{\rm m}^{\rm 40}({\rm exp})=3.361 \pm 0.075$~fm yield $r_{\rm m}^{\rm 48}=3.470 \pm 0.075$~fm, leading to 
$r_{\rm skin}^{\rm 48}=0.144 \pm 0.075$~fm.  
\item[Conclusion]
We conclude that 
$r_{\rm skin}^{208}({\rm exp})=0.324 \pm	(0.047)_{\rm exp} \pm (0.009)_{\rm th}~{\rm fm}$ for p scattering at $E_{\rm lab} = 534.1, 549, 806$~MeV. 
Our skin value  $r_{\rm skin}^{\rm 48}=0.144 \pm 0.083=0.061 \sim 0.227$~fm is consistent with $r_{\rm skin}^{48}({\rm CREX})=0.071 \sim 0.171$~fm. 
\end{description}
\end{abstract}

%\keywords{}

\maketitle

\clearpage

%\section{Introduction and conclusion}
%\label{Introduction}

{\bf Background:}
Horowitz {\it et al.} \cite{PRC.63.025501} proposed a direct measurement 
for neutron skin $r_{\rm skin}$. 
The measurement consists of parity-violating weak scattering and 
elastic electron scattering. 
The neutron radius $r_n$ is determined from the former experiment, whereas
the proton radius $r_p$ is from the latter. 

The direct measurement was applied for  $^{208}$Pb and $^{48}$Ca. 
As for $^{208}$Pb, the PREX collaboration presented 
\begin{equation}
r_{\rm skin}^{208}({\rm PREX2}) = 0.283\pm 0.071\,{\rm fm},
\end{equation}
combining the original Lead Radius EXperiment (PREX) 
result with the updated PREX2 result~\cite{Adhikari:2021phr,PRL.108.112502,PRC.85.032501}. 
As for $^{48}$Ca, the CREX group presented~\cite{CREX:2022kgg} 
\bea
r_{\rm skin}^{48}({\rm CREX})&=&0.121 \pm 0.026\ {\rm (exp)} \pm 0.024\ {\rm (model)}
\notag \\
&=&0.071 \sim 0.171~{\rm fm}. 
 \label{CREX-value}
 \eea
The $r_{\rm skin}^{208}({\rm PREX2}) $ and the $r_{\rm skin}^{48}({\rm CREX})$ are 
most reliable at the present 
stage, and provide crucial tests for the equation of state (EoS) of nuclear matter 
\cite{PRC.102.051303,AJ.891.148,AP.411.167992,EPJA.56.63,JPG.46.093003}
as well as nuclear structure.

Reed {\it et al.} \cite{arXiv.2101.03193} 
reported a value of the slope parameter of the EoS 
and examine the impact of such a stiff symmetry energy 
on some critical neutron-star observables.
The $r_{\rm skin}^{208}({\rm PREX2}) $ value 
is considerably larger than the other experimental 
values that are model-dependent~\cite{PRL.87.082501,PRC.82.044611,PRL.107.062502,PRL.112.242502}. 
Meanwhile, the nonlocal dispersive-optical-model 
(DOM) analysis of ${}^{208}{\rm Pb}$ yields  
$r_{\rm skin}^{\rm DOM} =0.25 \pm 0.05$ fm \cite{PRC.101.044303}. 
The value is consistent with $r_{\rm skin}^{208}({\rm PREX2})$.

Using the chiral (Kyushu) $g$-matrix folding model, we   
determine $r_{\rm skin}^{208}({\rm exp})=0.278 \pm 0.035$~fm from reaction cross section 
$\sigma_{\rm R}$ in $30 \leq E_{\rm lab} \leq 100$~MeV~\cite{Tagami:2020bee}. 
In addition, for  $^{4}$He+ $^{208}$Pb scattering, we determine  
$r_{\rm skin}^{208}({\rm exp})=0.416\pm 0.146$~fm  
from measured  $\sigma_R$ in $E_{\rm lab} = 30 \sim 50$ MeV~\cite{Matsuzaki:2021hdm}.
These values are consistent with $r_{\rm skin}^{208}({\rm PREX2})$.

For $^{12}$C scattering on  $^{9}$Be, $^{12}$C, $^{27}$Al targets, 
we tested reliability of the Kyushu $g$-matrix folding model  and found that 
the folding model is reliable in $30 \lsim E_{\rm lab} \lsim 100$~MeV and $250 \lsim E_{\rm lab} \lsim 400$~MeV~\cite{PRC.101.014620}. Furthermore, we mentioned that the difference between 
the $t$-matrix and  the $g$-matrix is small in $E_{\rm lab} \gsim 400$~MeV. Since the cutoff of the chiral nucleon-nucleon (NN) is 550~MeV, the chiral NN $t$-matrix is useful in 
$400 \lsim E_{\rm lab} \lsim 500$~MeV. For $E_{\rm lab} \gsim 400$~MeV, 
the most famous $t$-matrix is Love-Franey (LF) $t$-matrix~\cite{LF}.

As for $^{208}$Pb, it is possible to  determine reliable neutron radius $r_n({\rm  PREX2})=5.727 \pm 0.071$ fm and 
matter radius $r_m({\rm  PREX2})=5.617 \pm 0.044$~fm from $r_p({\rm exp})$ = 5.444 fm~\cite {PRC.90.067304} 
of electron scattering and  $r_{\rm skin}^{208}({\rm PREX2})$. 
The $r_p$ calculated with D1S-Gogny-HFB (D1S-GHFB) with the angular momentum  projection (AMP) agrees with $r_p({\rm exp})$. 
The neutron density calculated with D1S-GHFB+AMP is scaled so as to  $r_n^{\rm scaling}=5.727$ fm.
In Ref.~~\cite{WAKASA2021104749}, we showed that 
the  LF $t$-matrix folding model with the scaled neutron density and the D1S-GHFB+AMP proton 
one reproduces 
the data $\sigma_R({\rm exp})$~\cite{Dietrich:2002swm,Nakano:2021dau} 
at $E_{\rm lab} = 534.1, 549, 806$~MeV within total error bars. Nevertheless, we do not determine 
 $r_{\rm skin}^{208}$ from the data at $E_{\rm lab} = 534.1, 549, 806$~MeV.

As for $^{48}$Ca, an indirect measurement is made with the high-resolution $E1$ polarizability experiment (E1{\rm pE})~\cite{Birkhan:2016qkr}. 
The skin value $r_{\rm skin}^{48}(E1{\rm pE}) =0.14 \sim 0.20$~fm     
is consistent with  $r_{\rm skin}^{48}({\rm CREX})$.
Using  $^{4}$He+ $^{40}$Ca scattering in $E_{\rm lab} = 30 \sim 50$ MeV, 
we determine  matter radius $r_{\rm m}^{\rm 40}({\rm exp})$
from measured $\sigma_R$~\cite{Matsuzaki:2021hdm} ,  
whereas Zenihiro {\it el al. } deduce neutron radii $r_{\rm n}^{48,40}({\rm exp})$ from the angular
distributions of the cross sections and analyzing powers of polarized proton elastic scattering at 
$E_{\rm lab} = 295$~MeV~\cite{Zenihiro:2018rmz}. 
The $r_{\rm skin}^{48}({\rm exp})=0.168^{+0.025}_{-0.028}$~fm determined by Zenihiro {\it el al.} 
is consistent with $r_{\rm skin}^{48}({\rm CREX})$.

{\bf Aim:}
The first aim is to determine $r_{\rm skin}^{\rm 208}({\rm exp})$ 
from the data~\cite{Dietrich:2002swm,Nakano:2021dau} on $\sigma_{\rm R}$
of p+ $^{208}$Pb scattering at $E_{\rm lab} = 534.1, 549, 806$~MeV by 
using the  LF $t$-matrix folding model.

The second aim is to determine  $r_{\rm skin}^{\rm 48}({\rm exp})$ with 
the result $r_{\rm m}^{\rm 40}({\rm exp})=3.361 \pm 0.075$~fm~\cite{Matsuzaki:2021hdm} 
of $^{4}$He+$^{40}$Ca scattering in $E_{\rm lab} = 30 \sim 50$ MeV and 
the difference $\Delta \equiv r_{\rm m}^{48}({\rm exp})- r_{\rm m}^{40}({\rm exp})$, 
since there is no data on  $\sigma_{\rm R}$ for $^{4}$He+$^{48}$Ca scattering.
The derivation of $\Delta$ is shown below.

{\bf Method for determining $r_{\rm skin}^{\rm 48}({\rm exp})$:} 
Zenihiro {\it el al. } determine neutron radii $r_{\rm n}^{40}({\rm exp})=3.375^{+0.022}_{-0.023}$~fm and 
$r_{\rm n}^{48}({\rm exp})=3.555^{+0.025}_{-0.028}$~fm from the angular distributions of the cross sections and the analyzing powers of proton elastic scattering~\cite{Zenihiro:2018rmz}. 
We can obtain the proton radii for $^{40, 48}$Ca with  the isotope shift method based on the electron scattering~\cite{ADNDT.99.69}, i.e., $r_{\rm p}^{40}({\rm exp})=3.378$~fm and $r_{\rm p}^{48}({\rm exp})=3.385$~fm. 
Using these values, we can obtain   
$r_{\rm m}^{40}({\rm exp})=3.377^{+0.022}_{-0.023}$~fm, 
$r_{\rm m}^{48}({\rm exp})=3.485^{+0.025}_{-0.028}$~fm. 
%$r_{\rm skin}^{40}({\rm exp})=-0.003^{+0.022}_{-0.023}$~fm, 
%$r_{\rm skin}^{48}({\rm exp})=0.170^{+0.025}_{-0.028}$~fm. 

From the central values of $r_{\rm m}^{40}({\rm exp})$ and $r_{\rm m}^{48}({\rm exp})$, we obtain 
the difference $\Delta \equiv r_{\rm m}^{48}({\rm exp})- r_{\rm m}^{40}({\rm exp})$=0.109~fm. 
 In Ref.~\cite{Matsuzaki:2021hdm}, meanwhile, 
 we determined $r_{\rm m}^{\rm 40}({\rm exp})=3.361 \pm 0.075$~fm
from measured $\sigma_{\rm R}$ of $^{4}$He+ $^{40}$Ca scattering in  $E_{\rm lab} =30 \sim 50$~MeV. 
We can then obtain $r_{\rm m}^{48}({\rm exp})=3.470 \pm 0.075$~fm from 
$r_{\rm m}^{\rm 40}({\rm exp})=3.361 \pm 0.075$~fm and $\Delta$. 
The $r_{\rm m}^{48}({\rm exp})=3.470 \pm 0.075$~fm and  
$r_{\rm p}^{48}({\rm exp})=3.385$~fm lead to $r_{\rm skin}^{\rm 48}({\rm exp})=0.144 \pm 0.075$~fm, respectively.

{\bf Method for determining $r_{\rm skin}^{\rm 208}({\rm exp})$:} 
We use the folding model  based on Lovey-dovey (LF) $t$-matrix~\cite{LF} to determine 
$r_{\rm skin}^{\rm 208}({\rm exp})$ from data $\sigma_{\rm R}({\rm exp})$~\cite{Dietrich:2002swm,Nakano:2021dau} 
 at $E_{\rm lab} = 534.1, 549, 806$ MeV.
We have already applied the LF  $t$-matrix folding model for p+$^{4,6,8}$He scattering at 700~MeV to determine 
matter radii $r_m({\rm exp})$ from the high-accuracy data~\cite{Neumaier:2002eay}. 
The results  are $r_{m}({\rm exp})=2.48(3), 2.53(2)$~fm and $r_{\rm skin}=0.78(3), 0.82(2)$~fm 
for $^{6,8}$He~\cite{WAKASA2022105329}. 

 Now we show the formulation on  the LF  $t$-matrix folding model below. 
For proton-nucleus scattering, the potential $U(\vR)$ 
 between an incident proton  (p)  and a target (${\rm T}$) has the direct and exchange parts,
$U^{\rm DR}$ and $U^{\rm EX}$, as
\begin{subequations}
\begin{eqnarray}
U^{\rm DR}(\vR) & = & 
\sum_{\mu,\nu}\int             \rho^{\nu}_{\rm T}(\vrr_{\rm T})
            t^{\rm DR}_{\mu\nu}(s;\rho_{\mu\nu})  d
	    \vrr_{\rm T}\ ,\label{eq:UD} \\
U^{\rm EX}(\vR) & = & 
\sum_{\mu,\nu}
\int \rho^{\nu}_{\rm T}(\vrr_{\rm T},\vrr_{\rm T}+\vs) \nonumber \\
                &   &
\times t^{\rm EX}_{\mu\nu}(s;\rho_{\mu\nu}) \exp{[-i\vK(\vR) \cdot \vs/M]}
             d \vrr_{\rm T}\,~~
             \label{eq:UEX}
\end{eqnarray}
\end{subequations}
where $\vR$ is the relative coordinate between p  and T,
$\vs=-\vrr_{\rm T}+\vR$, and $\vrr_{\rm T}$ is
the coordinate of the interacting nucleon from T.
 Each of $\mu$ and $\nu$ denotes the $z$-component of isospin. 
The non-local $U^{\rm EX}$ has been localized in Eq.~\eqref{eq:UEX}
with the local semi-classical approximation~\cite{Brieva-Rook-1,Brieva-Rook-2,Brieva-Rook-3}
where $\vK(\vR)$ is the local momentum between p and T, 
and $M= A/(1 +A)$ for the mass number $A$ of T;
see Ref.~\cite{Minomo:2009ds} for the validity of the localization.

The direct and exchange parts, $t^{\rm DR}_{\mu\nu}$ and 
$t^{\rm EX}_{\mu\nu}$, of the $t$ matrix are described by
\begin{align}
t_{\mu\nu}^{\rm DR}(s) 
&=
\displaystyle{\frac{1}{4} \sum_S} \hat{S}^2 t_{\mu\nu}^{S1}
 (s) \hspace*{0.1cm}  \hspace*{0.1cm} 
 {\rm for} \hspace*{0.1cm} \mu+\nu = \pm 1,
 \\
t_{\mu\nu}^{\rm DR}(s) 
&=
\displaystyle{\frac{1}{8} \sum_{S,T}} 
\hat{S}^2 t_{\mu\nu}^{ST}(s) 
\hspace*{0.1cm}  \hspace*{0.1cm} 
{\rm for} \hspace*{0.1cm} \mu+\nu = 0,
\\
t_{\mu\nu}^{\rm EX}(s) 
&=
\displaystyle{\frac{1}{4} \sum_S} (-1)^{S+1} 
\hat{S}^2 t_{\mu\nu}^{S1} (s) 
\hspace*{0.1cm}  \hspace*{0.1cm} 
{\rm for} \hspace*{0.1cm} \mu+\nu = \pm 1, 
\\
t_{\mu\nu}^{\rm EX}(s) 
&=
\displaystyle{\frac{1}{8} \sum_{S,T}} (-1)^{S+T} 
\hat{S}^2 t_{\mu\nu}^{ST}(s) 
\hspace*{0.1cm}  \hspace*{0.1cm}
{\rm for} \hspace*{0.1cm} \mu+\nu = 0 
,
\end{align}
where $\hat{S} = {\sqrt {2S+1}}$ and $t_{\mu\nu}^{ST}$ are 
the spin-isospin components of the $t$-matrix interaction.

As proton and neutron densities, $\rho^{\nu=-1/2}_{\rm T}$ and $\rho^{\nu=1/2}_{\rm T}$, 
 we use D1S-GHFB+AMP; see Ref.~\cite{Tagami:2019svt} for the formulation.  
 As a way of taking the center-of-mass correction to the densities, 
we adapt the method of Ref.~\cite{Sumi:2012fr}.

We scale the D1S-GHFB+AMP neutron density so that  
the radius $r_n({\rm scaling})$ of the scaled density can reproduce $\sigma_{\rm R}({\rm exp})$, 
since the $r_p$ calculated with the  D1S-GHFB+AMP density agrees with 
$r_p^{\rm exp}=5.444$~fm~\cite{PRC.90.067304} of electron scaling. 
The same procedure is taken the D1M-GHFB+AMP neutron density, 
where D1M~\cite{Goriely:2009zz,Robledo:2018cdj} is an improved version of D1S and 
the proton radius calculated with D1M-GHFB+AMP agrees with $r_p^{\rm exp}=5.444$~fm.

Our scaling procedure is explained below. 
The scaled density $\rho_{\rm scaling}(\vrr)$ is determined  
from the original (D1S-GHFB+AMP or D1M-GHFB+AMP) one  $\rho(\vrr)$ as
\bea
\rho_{\rm scaling}(\vrr) \equiv \frac{1}{\a^3}\rho(\vrr/\a), ~\vrr_{\rm scaling} \equiv \vrr/\a
\label{eq:scaling}
\eea
with a scaling factor
\bea
\a=\sqrt{ \frac{\langle \vrr^2 \rangle_{\rm scaling}}{\langle \vrr^2 \rangle}} .
\eea
In Eq.~\eqref {eq:scaling}, we have replaced $\vrr$ by $\vrr/\a$ in the original density. 
Eventually, $\vrr$ dependence 
of   $\rho_{\rm scaling}(\vrr)$ is different from that of  $\rho(\vrr)$. 
We have multiplied the original density by $\a^{-3}$ 
 in order to normalize the scaled density. 
The symbol means $\sqrt{\langle \vrr^2 \rangle_{\rm scaling}}$ is the root-mean-square
radius of  $\rho_{\rm scaling}(\vrr)$.

\begin{figure}[H]
\begin{center}
\includegraphics[width=0.5\textwidth,clip]{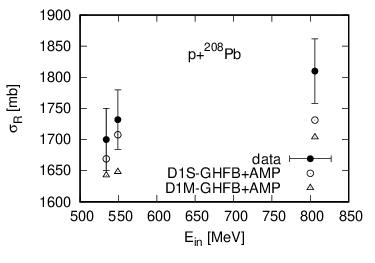}
 \caption{ 
 $E_{\rm lab}$ dependence of reaction cross sections $\sigma_{\rm R}$ 
 for $p$+$^{208}$Pb scattering. 
 Open circles stand for the results of the LF $t$-matrix  folding model with the D1S-GHFB+AMP densities, whereas open triangles correspond to that with the D1M-GHFB+AMP densities. 
 The data are taken from Refs.~\cite{Dietrich:2002swm,Nakano:2021dau}. 
  }
 \label{Fig-RXsec-p+Pb}
\end{center}
\end{figure}

 {\bf Results for  $r_{\rm skin}^{\rm 208}({\rm exp})$:} 
Figure~\ref{Fig-RXsec-p+Pb} shows   $\sigma_R $ as a function of $E_{\rm lab}$. 
The results of D1S-GHFB+AMP and  D1M-GHFB+AMP
are near the lower bound of data~\cite{Dietrich:2002swm,Nakano:2021dau}
in $500 \leq E_{\rm lab} \leq 900$~MeV.   
The result of D1S-GHFB+AMP is better than that of 
D1M-GHFB+AMP.

%%%%%%%%%%%%%%%%%%%%%%%
%%%  Figure
%%%%%%%%%%%%%%%%%%%%%%%
 
Now we scale the   D1S-GHFB+AMP neutron density so that 
the result of the LF $t$ matrix folding model agrees with the data~\cite{Dietrich:2002swm,Nakano:2021dau}. 
In the present case, the neutron scaling factor is $\a=1.017$. 
Since the resulting $r_n({\rm exp})$ depends on $E_{\rm lab}$, we take 
the the weighted mean and its total error for $E_{\rm lab} = 534.1, 549, 806$ MeV.
Neutron and matter radii thus obtained are $r_n({\rm exp})=5.768 \pm 0.047 $~fm and 
$r_m({\rm exp})=5.643 \pm 0.047 $~fm, leading to 
$r_{\rm skin}^{\rm 208}({\rm exp})=0.324 \pm 0.047 $~fm. 

The same procedure is taken for D1M-GHFB+AMP.
This leads to $r_{\rm skin}^{\rm 208}({\rm exp})=0.333 \pm 0.047 $~fm, where 
the neutron scaling factor is $\a=1.038$. The theoretical error 
is evaluated with the difference between the central values of D1S-GHFB+AMP and 
D1M-GHFB+AMP. The value is 0.009~fm. 
The result of D1S-GHFB+AMP yields better agreement with the data than that of 
D1M-GHFB+AMP.  We then obtain  
$r_{\rm skin}^{208}({\rm exp})=0.324 \pm	(0.047)_{\rm exp} \pm (0.009)_{\rm th}~{\rm fm}$.

{\bf Discussions:}
Finally, the uncertainties of our results are listed.
\begin{description}
   \item[~~~~~~~~~1.~Ambiguity of original densities taken]\mbox{}\\
As for proton and neutron densities for 
$^{48}$Ca, we used D1S and D1M in Ref.~\cite{TAGAMI2022105155} . 
Our result is  $r_{\rm skin}^{48}({\rm exp})=0.158 \pm	(0.023)_{\rm exp} \pm (0.012)_{\rm th}~{\rm fm}$; the theoretical error $(0.012)_{\rm th}~{\rm fm}$ is evaluated with D1S and D1M. 
The same procedure is taken for $^{208}$Pb. Our result is 
$r_{\rm skin}^{208}({\rm exp})=0.324 \pm	(0.047)_{\rm exp} \pm (0.009)_{\rm th}~{\rm fm}$.

  \item[~~~~~~~~~~2.~Experimental ambiguity]\mbox{}\\
Our present result $r_{\rm skin}^{48}=0.144 \pm 0.075$~fm based on $\Delta$
is consistent  with $r_{\rm skin}^{48}({\rm exp})=0.158 \pm	(0.023)_{\rm exp} \pm (0.012)_{\rm th}~{\rm fm}$ 
of Ref.~\cite{TAGAMI2022105155}.   
The central values are different from each other. The difference comes from the data used.

\end{description}

{\bf Conclusion:}
Our final values are 
$r_{\rm skin}^{208}({\rm exp})=0.324 \pm	(0.047)_{\rm exp} \pm (0.009)_{\rm th}~{\rm fm}$ and $r_{\rm skin}^{48}=0.144 \pm 0.075$~fm. 
Our values are consistent with $r_{\rm skin}^{208}({\rm PREX2}) $ and  
$r_{\rm skin}^{48}({\rm CREX}) $, respectively. 
These values are tabulated in Table \ref{skin values}.

%%%%%%%%%%%%%%
\begin{table}[htb]
\begin{center}
\caption
{Results for $r_{\rm skin}^{208}({\rm exp})$ and $r_{\rm skin}^{48}({\rm exp}) $. 
The values are shown in units of fm. 
 }
\begin{tabular}{cc}
\hline\hline
 & $r_{\rm skin}^{208}({\rm exp})$ or $r_{\rm skin}^{48}({\rm exp})$\\
\hline
 PREX2 & $0.283\pm 0.071$  \\
TW ($^{208}$Pb)& $0.324 \pm	(0.047)_{\rm exp} \pm (0.009)_{\rm th} $  \\
 CREX & $0.121 \pm 0.026{\rm (exp)} \pm 0.024{\rm (model)}$ \\
TW ($^{48}$Ca)  & $0.144 \pm 0.075 $  \\
\hline
\end{tabular}
 \label{skin values}
 \end{center} 
 \end{table}
%%%%%%%%%%%%%% 

\begin{acknowledgments}
We would like to thank Dr. Toyokawa from his contribution. 
\end{acknowledgments}

\bibliography{Folding-v11}

\end{document}